\documentclass[preprint,apjl]{aastex}
\usepackage{natbib}

\shorttitle{Pitch Angle of Galactic Spiral Arms}
\shortauthors{Michikoshi \& Kokubo}

\begin{document}

\title{
Pitch Angle of Galactic Spiral Arms 
}
\author{
Shugo Michikoshi\altaffilmark{1} and Eiichiro Kokubo\altaffilmark{2}
}
\altaffiltext{1}{
Department of Environmental Systems Science, Doshisha University, Tatara Miyakodani 1-3, Kyotanabe City, Kyoto 610-0394, Japan
}
\altaffiltext{2}{
Division of Theoretical Astronomy, National Astronomical Observatory of Japan, Osawa, Mitaka, Tokyo 181-8588, Japan
}
\email{smichiko@mail.doshisha.ac.jp and kokubo@th.nao.ac.jp}

\begin{abstract}

One of the key parameters that characterize spiral arms in disk galaxies is a pitch angle that measures the inclination of a spiral arm to the direction of galactic rotation. 
The pitch angle differs from galaxy to galaxy, which suggests that the rotation law of galactic disks determines it.  
In order to investigate the relation between the pitch angle of spiral arms and the shear rate of galactic differential rotation, we perform local $N$-body simulations of pure stellar disks.  
We find that the pitch angle increases with the epicycle frequency and decreases with the shear rate and obtain the fitting formula.   
This dependence is explained by the swing amplification mechanism.

\end{abstract}

\section{Introduction \label{sec:intro}}

Spiral structures are ubiquitous in various astrophysical disks.
In spiral galaxies, there are distinct spiral arm structures. 
Grand-design spiral galaxies have long continuous symmetric arms, while flocculent spiral galaxies have patchy irregular spiral arms.
In a protoplanetary gas disk, gravitational instability can occur during some phase of its evolution, which can produce spiral arms \citep{Gammie2001}. 
Also in Saturn's rings, the spiral structures in a broad sense exist.
The azimuthal brightness asymmetry is observed in the rings \citep[e.g.,][]{French2007}, 
which indicates the existence of the small scale spiral structures called as self-gravity wakes observed in $N$-body simulations \citep{Salo1992a, Salo1995}.
The wakes are caused by the gravitational instability of the ring. 

Our understanding of the origin of spiral arms in galaxies is still incomplete. 
One of the theories to explain spiral arms in galaxies is the density wave theory \citep{Lin1964, Lin1966}. 
Spiral structures are considered as a quasi-stationary standing wave pattern that rotates around the galactic center with a constant pattern speed. 
The spiral arms may be excited by tidal interactions with companion galaxies \citep[e.g.,][]{Oh2008} or the central bars \citep[e.g.,][]{Buta2005, Salo2010}. 

In a differentially rotating disk, a leading density pattern rotates to a trailing one due to the shear. 
If Toomre's $Q$ value is larger than unity but not too much, the amplitude of the pattern can be enhanced during the rotation.
This mechanism is called swing amplification \citep{Goldreich1965, Julian1966, Toomre1981}. 
If a perturber such as the corotating over-dense region exists, trailing patterns form \citep{Julian1966}. 
In $N$-body simulations, since a disk consists of a finite number of stars, small density noise always exists.
Thus, even if there is not a perturber, the small leading wave always exists, and the trailing wave can grow spontaneously due to the swing amplification mechanism \citep{Toomre1991}. 
The spirals generated by the swing amplification are not stationary but transient and recurrent, which appear and disappear continuously.
This transient and recurrent picture is supported by $N$-body simulations for multi-arm spirals \citep{Sellwood1984, Baba2009, Sellwood2000, Sellwood2010, Fujii2011}. 

The linear theory of the swing amplification gives the amplification factor and the most unstable wavelength, but it cannot explain the overall evolution of spiral arms.
\cite{Baba2013} studied the dynamics of stars in spiral arms and found that the nonlinear particle wave interaction is important to understand the damping and growing phase of spiral arms.
\cite{DOnghia2013} performed high-resolution $N$-body simulations including initial density inhomogeneities that induce the spiral patterns due to the swing amplification. 
Once spiral arms form, the spiral arms remain.
This results from the nonlinear effect.
The local underdense and overdense regions act as perturbers, which maintain the spiral structure.

One of the key parameters to characterize the morphology of spiral galaxies is the pitch angle. 
The pitch angle is the angle between the tangents to a spiral arm and a perfect circle, which measures how tightly the spiral arms are wound. 
\cite{Julian1966} investigated the response of the particle density to an imposed perturbation using the collisionless Boltzmann equation. 
They found the trend that the pitch angle decreases with the shear rate.
The correlation between the shear rate and the pitch angle enables us to determine a rotation curve from the spiral structure.

The epicycle frequency $\kappa$ is related to the shear rate $\Gamma$:
\begin{equation}
 \Gamma = \frac{2 A}{\Omega} = 
 - \frac{\mathrm{d}\log\Omega}{\mathrm{d}\log R} =
 2 - \frac{\kappa^2}{2 \Omega^2},
 \label{eq:sheartate}
\end{equation}
 where $A$ is the first Oort constant, and $\Omega$ is the circular frequency. 
The observational study shows the relation that the pitch angle
 decreases with the shear rate, and the fitting formula is given as
 \citep{Seigar2005, Seigar2006}: 
\begin{equation}
  \theta = (64.25 \pm 2.87)^\circ - \Gamma (36.62 \pm 2.77) ^\circ .
  \label{eq:seigar}
\end{equation}

\cite{Grand2013} performed global $N$-body simulations and investigated the spiral patterns using Fourier analysis. 
From the spiral phase variation they calculated the pitch angle of the spiral arm.  
They found that galaxies of the higher shear rate have the  smaller pitch angle. 
They did not study the dependence of the pitch angle on Toomre's $Q$ value since $Q$
 evolves over time. 
It is expected that the pitch angle barely depends on $Q$ from
 \cite{Julian1966}. 

In order to understand the dynamics of spiral arms, we investigate the
 pitch angle dependence on the shear rate by local $N$-body simulations of pure stellar disks. 
Section \ref{sec:simulation} summarizes the calculation method.
In Section \ref{sec:result}, we present the simulation results.
In Section \ref{sec:discussion}, we discuss the relation between the
 pitch angle and the shear rate by using the linear theory. 
Section \ref{sec:summary} gives a summary.

\section{Calculation Method \label{sec:simulation}}
\subsection{Model}

We perform local $N$-body simulations of pure stellar disks based on the epicycle approximation.
We do not consider an entire disk but a small rotating patch by employing a local shearing box 
 \citep[e.g.,][]{Toomre1991, Fuchs2005}.  This treatment reduces the number of necessary particles in a simulation
 significantly, and enables us to perform high resolution simulations. 
We consider a small patch of a disk such that $L_x, L_y \ll r$,
 where $L_x$ and $L_y$ are the width and length of the 
 patch and $r$ is the galactocentric distance of the patch.
We adopt a local Cartesian coordinate system ($x,y,z$), whose origin
 revolves around the galactic center with the circular frequency $\Omega$, which is given by
\begin{equation}
 \Omega^2 =\frac{1}{r} \left( \frac{\partial \Phi}{\partial r} \right),  
\end{equation}
where $\Phi$ is the axisymmetric galactic potential.
The $x$-axis is directed radially outward, the $y$-axis is parallel to
 the direction of rotation, and the $z$-axis is normal to the $x$-$y$
 plane.  
In the epicycle approximation, neglecting the higher order terms with respect to $x$, $y$, and $z$, the equation of motion of particle $i$ is given by 
\begin{eqnarray}
  \frac{\mathrm d^2 x_i}{\mathrm{d}t^2} &=& 2 \Omega \frac{\mathrm{d} y_i}{\mathrm{d}t} + \left(4 \Omega^2 - \kappa^2 \right) x_i + \sum_{j \ne i} \frac{G m (x_j - x_i)}{(r_{ij}^2+\epsilon^2)^{3/2}}, \nonumber  \\
  \frac{\mathrm d^2 y_i}{\mathrm{d}t^2} &=& - 2 \Omega \frac{\mathrm d x_i}{\mathrm{d}t} + \sum_{j \ne i} \frac{G m (y_j - y_i)}{(r_{ij}^2+\epsilon^2)^{3/2}},  \label{eq:eom} \\
  \frac{\mathrm d^2 z_i}{\mathrm{d}t^2} &=& - \nu^2 z_i + \sum_{j \ne i} \frac{G m (z_j - z_i)}{(r_{ij}^2+\epsilon^2)^{3/2}}, \nonumber
\end{eqnarray}
 where $r_{ij}$ is the distance between particles $i$ and $j$, $m$ is the particle mass
 \citep[e.g.,][]{Toomre1981, Toomre1991, Kokubo1992, Fuchs2005}. 
In Equation (\ref{eq:eom}), $2 \Omega \mathrm{d} y_i/\mathrm{d}t$ and $- 2 \Omega \mathrm d x_i/\mathrm{d}t $ are Coriolis force, $4 \Omega^2 x_i$ is the centrifugal force, $-\kappa^2 x_i$ and $-\nu^2 z_i$ are the galactic gravitational force, and the terms proportional to $(r^2_{ij} + \epsilon^2)^{-3/2}$ are the gravitational force from the other particles.
We assume that all particles have the same mass. 
The length $\epsilon$ is the softening parameter 
 $\epsilon = r_\mathrm{t}/4$ where $r_\mathrm{t}$ is the tidal radius of
 a particle:  
\begin{equation}
  r_\mathrm{t} = \left(\frac{2m G}{4 \Omega^2-\kappa^2} \right)^{1/3}.
\end{equation}
The frequencies $\kappa$ and $\nu$ are the epicycle and vertical frequencies at the center of the computational box:
\begin{eqnarray}
\kappa^2 &=& \frac{1}{r} \left( \frac{\partial \Phi}{\partial r} \right) + 4 \Omega^2, \label{eq:kappa} \\
\nu^2 &=& \frac{\partial^2 \Phi}{\partial z^2} \label{eq:nu} .
\end{eqnarray}
Since the size of the computational box is small, we can assume that all particles in the computational box have the same $\kappa$ and $\nu$.

The motion of particles is pursued only in the computational box with the periodic boundary condition \citep{Wisdom1988}.
There are copied boxes around the computational box.
When a particle in the computational box crosses the boundary, the corresponding particle in the copied box comes into the computational box through the opposite boundary. 
The position and velocity of the particle that crosses the boundary is calculated by considering the velocity shear.

The size of the computational box $L_x$ and $L_y$ should be sufficiently larger than the characteristic scale of the spiral arms  
that is the critical wavelength of the gravitational instability
\begin{equation}
  \lambda_\mathrm{cr} = \frac{4 \pi^2 G \Sigma_0}{\kappa^2},
\end{equation}
where $\Sigma_0$ is the initial surface density.
We set the size of the computational box as $L_x = L_y = L = 5 \lambda_\mathrm{cr}$.

We set the unit time as $\Omega^{-1}$ and the unit length as $r_\mathrm{t}$ \citep{Kokubo1992}.
The equation of motion is integrated using a second-order leapfrog integrator with time-step $\Delta t = (2 \pi / \Omega)/200$. 
We calculate the self-gravity of particles not only in the computational box but also from the surrounding copied boxes.  
The cutoff length of the gravity is $L_\mathrm{cut} = \mathrm{min}(L_x , L_y)$.
The self-gravity of particles, which is the most computationally expensive part, is calculated using the special-purpose computer, GRAPE-7 \citep{Kawai2006}.

\subsection{Initial Conditions}

We assume that the initial surface density $\Sigma_0$ of particles in the computational box is uniform.
The total mass in area $\lambda_\mathrm{cr}^2$ is fixed and the
 particle mass is given by 
 $m = \lambda_\mathrm{cr}^2 \Sigma_0/N_\mathrm{c}$ 
 where $N_\mathrm{c}$ is the number of particles in $\lambda_\mathrm{cr}^2$. 
We set $N_\mathrm{c}=8000$ and then the total number of particles is
 $N=N_\mathrm{c} L_x L_y/\lambda_\mathrm{cr}^2=2.0 \times 10^5$. 
If we neglect the weak dependence of the Coulomb logarithm on
 $N_\mathrm{c}$ and assume $\log \Lambda \simeq 5$, the two-body
 relaxation time is proportional to $N_\mathrm{c}$, which is estimated as \citep[e.g.,][]{Kokubo1992}
 \begin{equation}
   t_\mathrm{r} \simeq 2 \times 10^2 (Q/1.4)^4 (N_\mathrm{c}/8000) \Omega^{-1}.
 \end{equation}
Since the simulation time is much shorter than the relaxation time, the two-body relaxation barely affects the dynamical evolution.

The initial Toomre's $Q$ value is 
\begin{equation}
  Q_\mathrm{ini}= \frac{\sigma_x \kappa}{3.36 G \Sigma_0},
\end{equation}
where $\sigma_x$ is the initial radial velocity dispersion \citep{Toomre1964}.  
The initial radial velocity dispersion $\sigma_x$ is calculated from $Q_\mathrm{ini}$.
We adopt the triaxial Gaussian model as the velocity distribution.
In the epicycle approximation, the ratio of azimuthal to radial velocity dispersions is $\sigma_y / \sigma_x = \kappa / 2 \Omega $ \citep[e.g.,][]{Binney2008}.
The ratio of the radial to vertical velocity dispersions  $\sigma_z / \sigma_x$ depends on $\kappa$ and $\sigma_z$.
The ratio $\sigma_z / \sigma_x$ increases with $\kappa$.
For $\sigma_z \lesssim r_\mathrm{t} \Omega$, the ratio is $\sigma_z / \sigma_x \sim 0.5$--$0.8$ \citep{Ida1993}.  We adopt the simple linear model $\sigma_z/ \sigma_x = 0.3 \kappa / \Omega + 0.2$ for $\sigma_z \lesssim r_\mathrm{t} \Omega$.
The vertical distribution of particles is determined so that it is consistent with the velocity distribution, and $x$ and $y$ of particles are distributed randomly.

There are 8 parameters $\kappa$, $\nu$, $Q_\mathrm{ini}$, $N_\mathrm{c}$, $L$,  $L_\mathrm{cut}$, $\epsilon$, and $\Delta t$ in the simulation model.
We mainly explore the two parameters, $\tilde \kappa = \kappa / \Omega$ and $Q_\mathrm{ini}$ .
We have 50 simulation models (1a--1j, 2a--2j, 3a--3j, 4a--4j, 5a--5j),
 where  $Q_\mathrm{ini} = 1.0$(1), $1.2$(2), $\ldots$, and $1.8$(5), and
 $\tilde \kappa= 1.0$ (a), $1.1$ (b), $\ldots$, and $1.9$ (j), respectively. 
We have checked that the following results barely depend on the other parameters $N_\mathrm{c}$, $L$, $L_\mathrm{cut}$, $\epsilon$, and $\Delta t$. 
We adopt the vertical frequency $\nu=3 \Omega$.
We also have performed the other 50 simulation models with $\nu=\Omega$ and confirmed that the following results barely depend on $\nu$.

\section{Pitch Angle \label{sec:result}}
\subsection{Spatial Correlation}
\begin{figure}

 \begin{minipage}{0.5\hsize}
  \begin{center}
   \includegraphics[width=0.9\columnwidth]{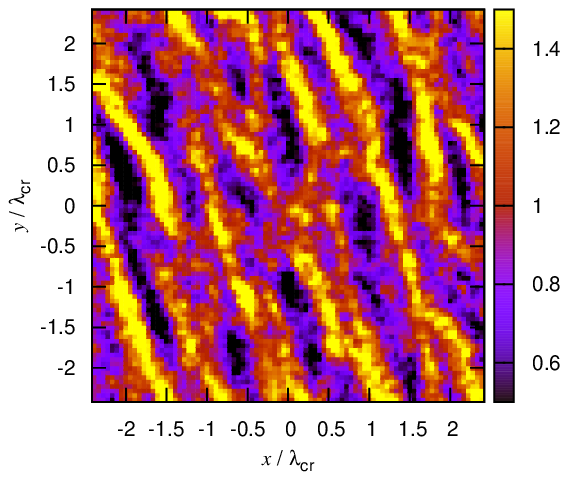} 
   \hspace{1.0cm} (a) 
  \end{center}
 \end{minipage}
 \begin{minipage}{0.5\hsize}
  \begin{center}
   \includegraphics[width=0.9\columnwidth]{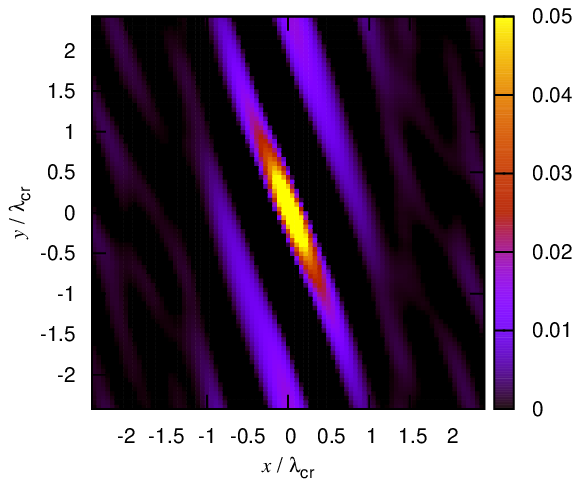} 
   \hspace{1.0cm} (b) 
  \end{center}
 \end{minipage}

 \hspace{1.6cm}  

 \begin{minipage}{0.5\hsize}
  \begin{center}
   \includegraphics[width=0.9\columnwidth]{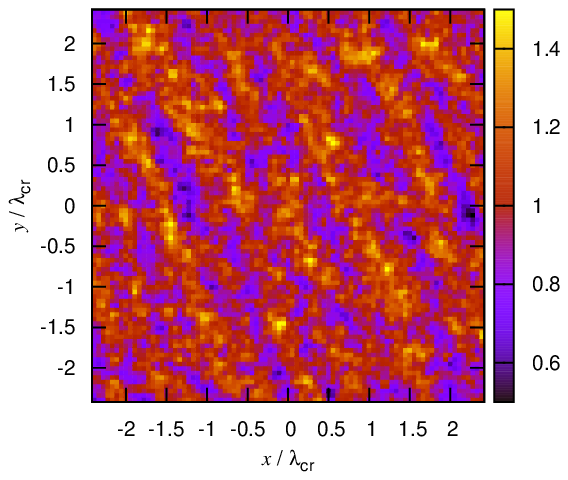} 
   \hspace{1.0cm} (c) 
  \end{center}
 \end{minipage}
 \begin{minipage}{0.5\hsize}
  \begin{center}
   \includegraphics[width=0.9\columnwidth]{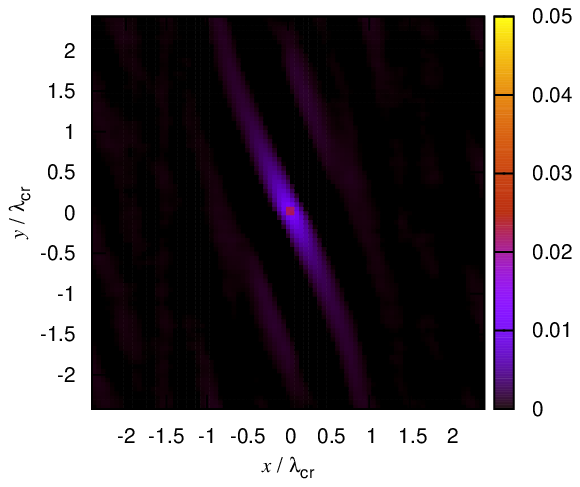} 
   \hspace{1.0cm} (d) 
  \end{center}
 \end{minipage}

 \hspace{1.6cm}  

 \begin{minipage}{0.5\hsize}
  \begin{center}
   \includegraphics[width=0.9\columnwidth]{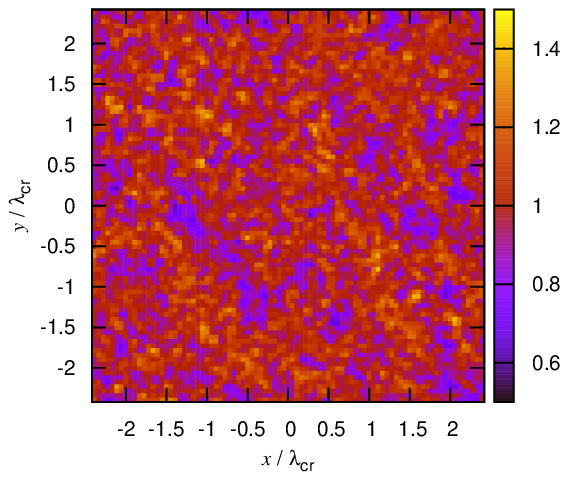} 
   \hspace{1.0cm} (e) 
  \end{center}
 \end{minipage}
 \begin{minipage}{0.5\hsize}
  \begin{center}
   \includegraphics[width=0.9\columnwidth]{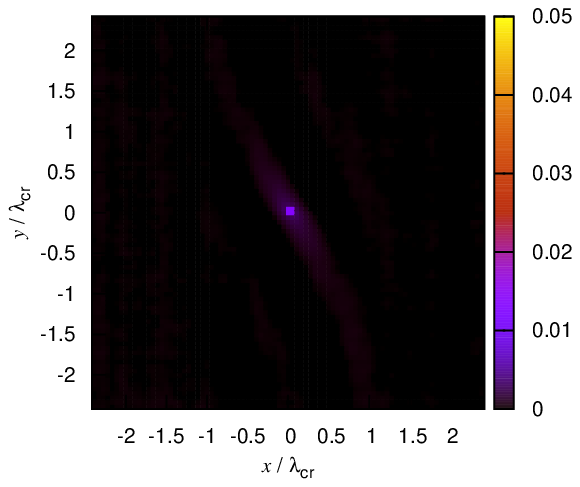} 
   \hspace{1.0cm} (f) 
  \end{center}
 \end{minipage}

 \caption{
The surface density distribution in the $x$-$y$ plane (left panels)
 at $t=2.0 \times 2 \pi / \Omega$ and the time-averaged spatial
 correlation (right panels) for $Q=1.0$ (model 1e) (top panels) and
 $Q=1.4$ (model 3e) (middle panels) and $Q=1.8$ (model 5e) (bottom
 panels). 
The epicycle frequency of all models is $\tilde \kappa=1.4$.
The surface density is normalized by the average initial surface density.
 }
\label{fig:snap}
\end{figure}

In order to investigate the pitch angle quantitatively, we calculate the spatial correlation function $\xi$: 
\begin{equation}
  \xi(x,y) = -1 + \frac{1}{\Sigma_0^2 L^2} \int \!\!\! \int _{-L/2}^{L/2} \Sigma(x+x', y+y') \Sigma(x',y') \mathrm{d}x' \mathrm{d}y'.
\end{equation}
We calculate the surface density with the uniform grid of $90 \times 90$. 
Figure \ref{fig:snap}a shows the particle surface density distribution at $t=2.0 \times 2 \pi / \Omega$ for model 1a where $\tilde{\kappa} = 1.4$ and $Q_\mathrm{ini} = 1.0$.  
Spiral or wake structures are formed due to gravitational instability. They are trailing, that is, the pitch angle is positive. 
Figure \ref{fig:snap}b shows the time-averaged $\xi$ over $3 \times 2\pi /\Omega$.
The most prominent feature is the inclined straight line crossing the center, in other words, a trailing pattern.  
For $Q_\mathrm{ini}=1.4$, the basic features are the same as those for $Q_\mathrm{ini}=1.0$.
Figures \ref{fig:snap}c and \ref{fig:snap}d show the clear trailing patterns.
However their amplitude is smaller than those for $Q_\mathrm{ini}=1.0$.
This is because the amplification factor of the swing amplification decreases with Toomre's Q value \citep{Toomre1981}.
For $Q_\mathrm{ini}=1.8$, the wakes are trailing but faint and thus the spatial correlation is very weak (Figure \ref{fig:snap}e and \ref{fig:snap}f). 
The models of $Q_\mathrm{ini}=1.6$ show the similar tendency to those of
 $Q_\mathrm{ini}=1.8$. 
The distinct spirals do not form for $Q_\mathrm{ini} \gtrsim 1.5$. 

We measure the pitch angle from the spatial correlation.
The pitch angle is the angle between the vertical line and the
 correlation ridge that is approximated by the straight line crossing
 the origin. 
We define the pitch angle of spirals as the angle $\theta$ where the
 function $f(\theta)$ has the maximum value, where $f(\theta)$ is
 \citep{Wakita2008} 
\begin{equation}
 f(\theta) = 
 \int_{-L/2}^{L/2} \xi(s \sin \theta, -s \cos \theta) \mathrm{d}s.
\end{equation}

\begin{figure}
 \begin{center}
  \plotone{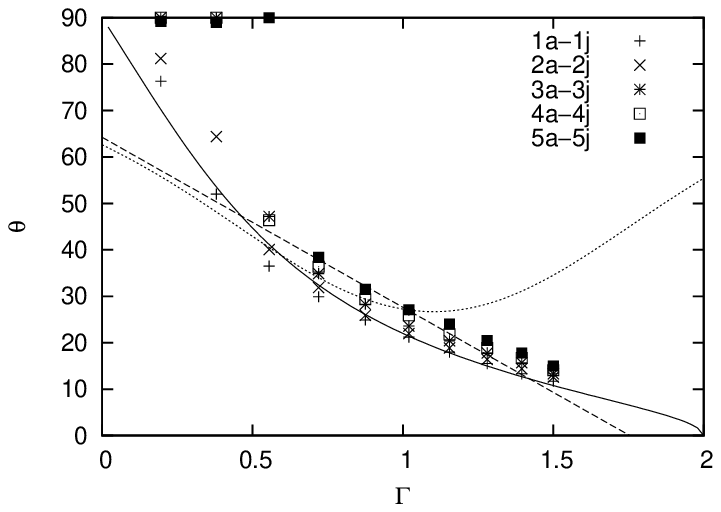}
 \end{center}
 \caption{
The pitch angle $\theta$ as a function of $\Gamma$ for the initial $Q$ values  $1.0$, $1.2$, $1.4$, $1.6$, and $1.8$. 
The solid curve shows the fitting formula described by Equation
 (\ref{eq:esti}), and the dashed curve corresponds to the observational
 fitting given by Equation (\ref{eq:seigar}) \citep{Seigar2006}. 
The dotted line is the formula proposed by the linear theory for $0.2 \le \Gamma \le 1.0$ \citep{Fuchs2001, Baba2013}. 
}
\label{fig:kappatheta}
\end{figure}

The pitch angle dependence on $\Gamma$ and $Q_\mathrm{ini}$ is
 shown in Figure \ref{fig:kappatheta}.  
The pitch angle decreases with the shear rate $\Gamma$.
For the small shear rate, since the winding due to the shear is weak, the pitch angle is large. 
The pitch angle increases with $Q_\mathrm{ini}$, but its dependence is very weak.  
The shear rate $\Gamma$ is more important than $Q_\mathrm{ini}$. 
The dashed curve in Figure \ref{fig:kappatheta} is calculated from the
 observational fitting formula of Equation (\ref{eq:seigar})
 \citep{Seigar2006}. 
Roughly speaking, Equation (\ref{eq:seigar}) agrees with the simulation
 results.  
However, the fitting values for $0.6 < \Gamma < 1.2$ are larger
 than those from the simulations systematically.  
Furthermore, the observational fitting formula is the linear function of $\Gamma$, but as shown in Figure \ref{fig:kappatheta}, it
 seems that the pitch angle is a convex function of $\Gamma$. 

\cite{Fuchs2001} derived an empirical formula of the azimuthal wavenumber of the most amplified wave. 
\cite{Baba2013} used this empirical formula and assumed that the radial wavelength is equal to the critical wavelength and derived the pitch angle:
\begin{equation}
\tan \theta = 1.932 - 5.186 \left(\frac{\Gamma}{2}\right) + 4.704 \left(\frac{\Gamma}{2}\right)^2.
\end{equation}
As shown in Figure \ref{fig:kappatheta}, this pitch angle formula agrees with our results for $0.2 \le \Gamma \le 1.0$, but does not for $\Gamma<0.2$ or $1.0< \Gamma$.
This is mainly caused by the limitation of the fitting formula of the azimuthal wave number of  \cite{Fuchs2001}, which is applicable only for $0.2 \le \Gamma \le 1.0$.
In addition, strictly speaking, the radial wavelength can be different from the critical wavelength and depends on the shear rate $\Gamma$.

We derive a new formula from the results of the numerical simulations.
If we neglect any interactions among particles, the spiral arm swings
 from leading to trailing due to differential rotation, and the pitch
 angle evolution is described as \citep[e.g.,][]{Binney2008} 
\begin{equation}
  \tan \theta = \frac{1}{2At}.
  \label{eq:pitc}
\end{equation}
If we choose about half an epicycle period $t \simeq 3.5 / \kappa$, from Equation (\ref{eq:pitc}) we obtain 
\begin{equation}
 \tan \theta \simeq \frac{1}{7} \frac{\kappa}{A} = 
 \frac{4}{7} \frac{\tilde \kappa}{4 - \tilde \kappa^2} =
 \frac{2}{7} \frac{\sqrt{4-2\Gamma}}{\Gamma}. 
 \label{eq:esti}
\end{equation}
The solid curve in Figure \ref{fig:kappatheta} corresponds to the pitch angle given by Equation (\ref{eq:esti}), which agrees well with the results of the simulations. 

It is not trivial that Equation (\ref{eq:pitc}) with $t \simeq 3.5 / \kappa$ gives the pitch angle of spiral arms.
In Section \ref{sec:discussion}, we discuss the derivation of the pitch angle formula from the linear analysis.

\subsection{Fourier Transformation}
We can extract the dominant wave mode using the Fourier analysis.
The Fourier transformation of the surface density is defined by 
\begin{equation}
\hat \Sigma (k_x, k_y) = \int \!\!\! \int \Sigma(x,y) \exp(i (k_x x + k_y y)) \mathrm{d}k_x \mathrm{d}k_y,
\end{equation}
where $k_x$ and $k_y$ are the radial and azimuthal wavenumbers.

\begin{figure}
   \includegraphics[width=0.33\columnwidth]{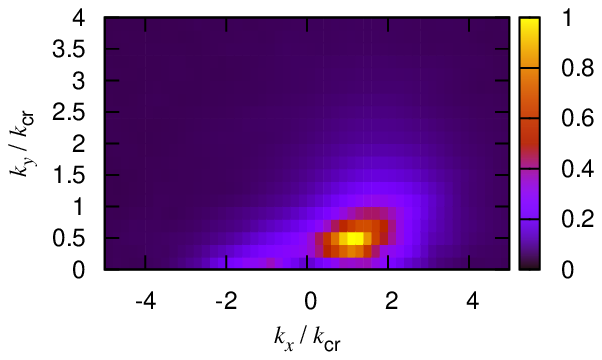}
   \includegraphics[width=0.33\columnwidth]{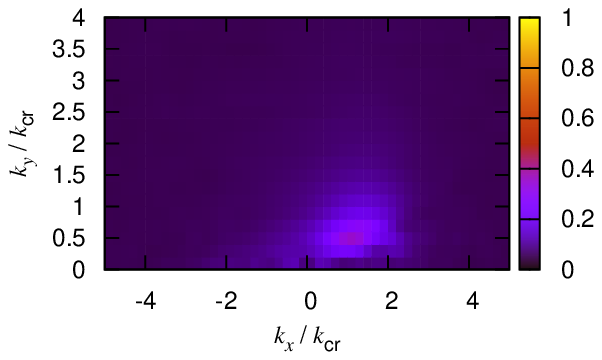}
   \includegraphics[width=0.33\columnwidth]{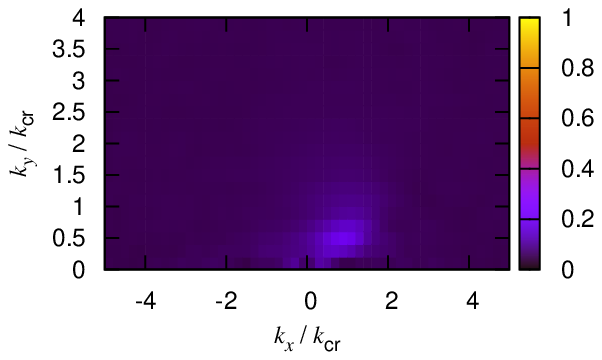}
 \caption{
   The time-averaged Fourier amplitude $| \hat \Sigma|$ for $\tilde \kappa=1.4$, $Q_\mathrm{ini} = 1.0, 1.4, 1.8$ (models 1e, 3e, 5e) (left, middle, right panels, respectively).
   The wavenumber is normalized by $k_\mathrm{cr} = 2 \pi / \lambda_\mathrm{cr}$.
}
\label{fig:fourier}
\end{figure}

Figure \ref{fig:fourier} shows the time-averaged Fourier amplitude over $3 \times \Omega/2\pi$ for $\tilde \kappa = 1.4$ and $Q_\mathrm{ini} = 1.0, 1.4,$ and  $1.8$ (models 1e, 3e, and 5e).
In these models the Fourier amplitude has the maximum at $ (k_x, k_y) \simeq (1.0 k_\mathrm{cr}, 0.5 k_\mathrm{cr})$.
The wavenumber of the dominant mode does not depend on $Q_\mathrm{ini}$.
However, the amplitude of the wave depends on $Q_\mathrm{ini}$.
As $Q_\mathrm{ini}$ increases, the maximum amplitude decreases.
For large $Q_\mathrm{ini}$, the peak position is obscure.

\begin{figure}
   \plotone{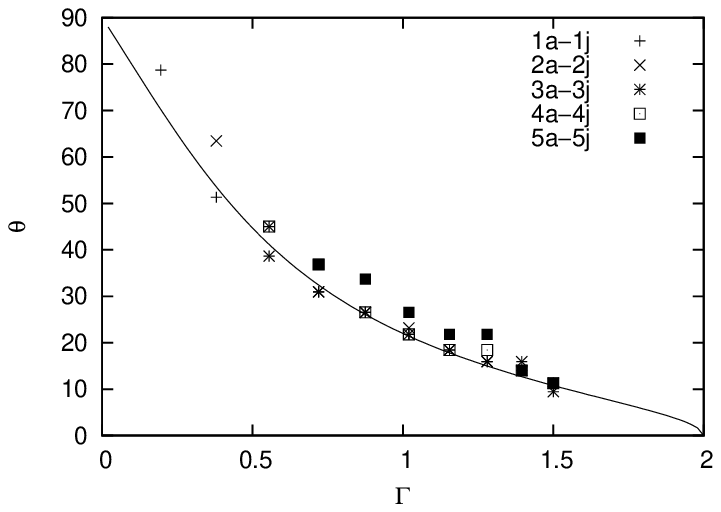}
 \caption{
   The same as Figure \ref{fig:kappatheta} but the pitch angle is calculated from the Fourier transformation.
}
\label{fig:pitchangle_f}
\end{figure}

The pitch angle of the wave with $(k_x, k_y)$ is 
\begin{equation}
\tan \theta = \frac{k_y}{k_x}.
\label{eq:fft_theta}
\end{equation}
The spiral arm corresponds to the dominant wave whose amplitude is the maximum.
We can calculate the pitch angle of the spiral arm from Equation (\ref{eq:fft_theta}) using the wavenumber of the dominant wave.
We compare the pitch angle from the Fourier transformation with that from the correlation function. 
Figure \ref{fig:pitchangle_f} shows the pitch angle from the Fourier transformation.
For $Q_\mathrm{ini}< 1.5$, the pitch angle from the Fourier transformation is the same as those from the spatial correlation.

However, if $Q_\mathrm{ini} = 1.6,$ and $1.8$ and $\Gamma<1.0$, we can see the difference of the pitch angle.
For $\Gamma<0.4$, we cannot obtain the pitch angle of the trailing wave because the amplitude of the wave is too small to extract the dominant mode for $Q_\mathrm{ini} = 1.6$ and $1.8$.
The extraction of the dominant wave mode by the Fourier analysis fails.
In these parameters, the correlation method gives the more accurate pitch angle than the Fourier analysis.

\section{Linear Analysis \label{sec:discussion}}
In the swing amplification mechanism, while the wavelet rotates from leading to trailing due to the shear, the wavelet is amplified.
Thus, the initial leading wavelet is necessary.
Since the number of particles is finite, the Poisson noise implies the leading mode that has the small amplitude.
After the first spiral arms are formed, the activity of the rapid spiral formation and destruction continues.
This indicates that the leading mode is always generated. 
We do not discuss the origin of the leading mode here, but it may be generated by some nonlinear processes \citep[e.g.,][]{Fuchs2005, DOnghia2013}.
The overall evolution of spiral arms cannot be obtained by the linear theory.
However, the linear theory can often capture some aspects of the basic physics.
If we assume that the spiral arm corresponds to the most amplified wave that is predicted by the linear theory, it is expected that the shape of spiral arms can be explained by the linear theory.

We investigate the pitch angle dependence on the shear rate using the linear theory \citep{Julian1966}. 
We focus on a single wavelet with  $k_x$, $k_y$, and density amplitude $D$. 
Due to the shear, the normalized radial wavenumber $\tilde k_x= k_x /(2 \pi/\lambda_\mathrm{cr}) $ increases with time $\tilde t$
\begin{equation}
  \tilde k_x (\tilde t) = \frac{\Gamma \tilde k_y }{\tilde \kappa} \tilde t, \label{eq:kx}
\end{equation}
where $\tilde t$ is the normalized time $\tilde t = t \kappa$, and $\tilde k_y$ is the normalized azimuthal wavenumber $\tilde k_{y} = k_y / ( 2 \pi /\lambda_\mathrm{cr})$ that is the inverse of $X$ in \cite{Julian1966}: $\tilde k_y = 1/X$, while $\tilde k_y$ is constant.
The wavelet is trailing when $\tilde t>0$ ($\tilde k_x>0$) and leading when $\tilde t<0$ ($\tilde k_x<0$). 

As the wavelet rotates, the density amplitude $D$ varies with $\tilde t$.
The density amplitude evolution is given by the integral equation
 \citep{Julian1966}: 
\begin{equation}
 D(\tilde t) = 
 \int_{\tilde t_\mathrm{i}}^{\tilde t} 
 K(\tilde t',\tilde t;\kappa,  Q, \tilde k_{y}) 
 (D_{\mathrm{imp}} + D(\tilde t')) \mathrm{d} \tilde t', 
 \label{eq:int}
\end{equation}
 where $K$ is the kernel function, and $D_{\mathrm{imp}}$ is the density amplitude by the imposed perturbation.

We consider the wavelet excited at the initial time $\tilde t = \tilde t_\mathrm{i}$ due to some disturbance, and neglect any disturbance to the wavelet after $\tilde t = \tilde t_\mathrm{i}$, that is, we assume $D_{\mathrm{imp}}=0$ for $t>t_\mathrm{i}$.
From Equation (\ref{eq:kx}), $\tilde t_\mathrm{i}$ is related to the
initial radial wave number  $\tilde k_{x\mathrm{i}} = \Gamma \tilde k_y \tilde t_\mathrm{i}/ \tilde \kappa $. 
Therefore, the solution to the integral equation $D(\tilde t)$ depends on the four
 dimensionless parameters $\tilde \kappa$, $Q$, $\tilde k_{x\mathrm{i}}$ and $\tilde k_y$. 
The two parameters $\tilde \kappa$ and $Q$ stand for a disk model, and
 the other two parameters $\tilde k_{x\mathrm{i}}$ and $\tilde k_y$
 specify the wavelet that we focus on. 

\begin{figure}
 \begin{center}
  \plotone{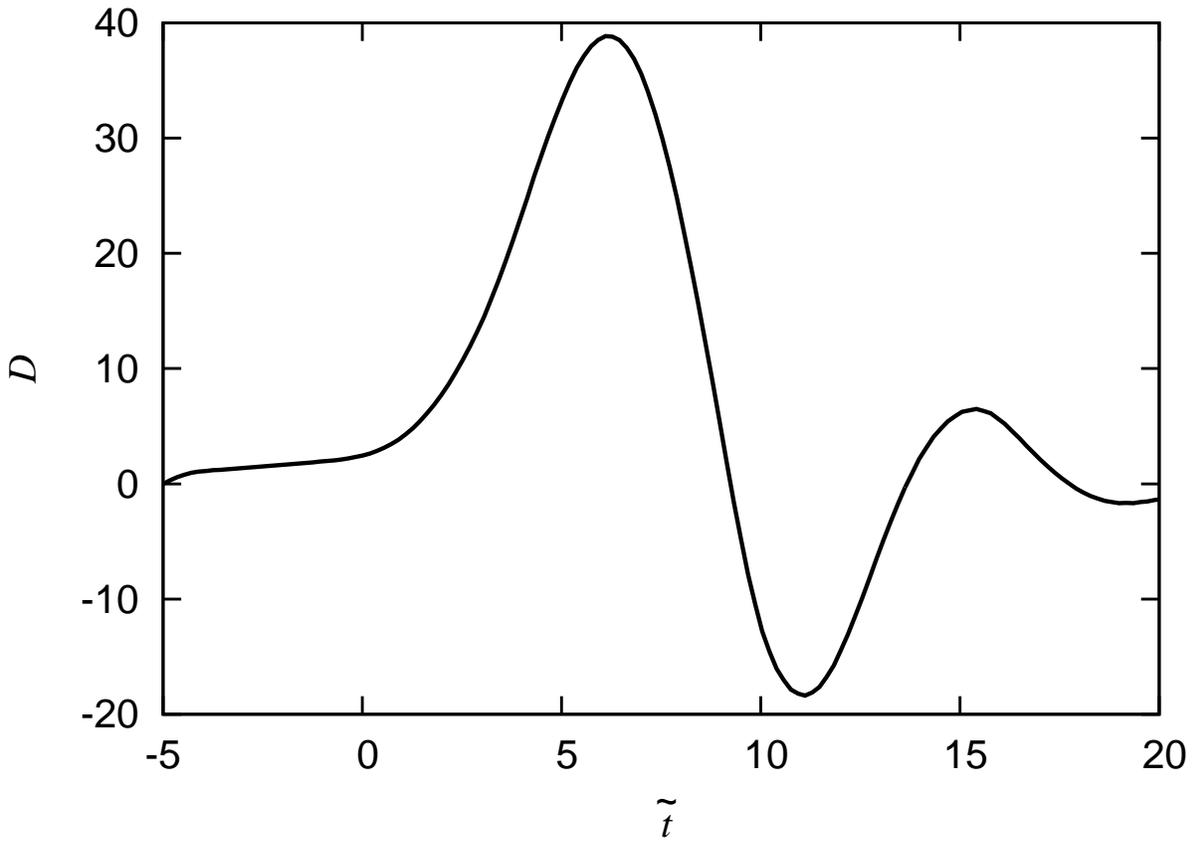}
 \end{center}
 \caption{
The time evolution of the wavelet amplitude $D$ calculated by the linear
 theory  for $\tilde \kappa=1.4$, $Q=1.2$, $\tilde k_{x\mathrm{i}} = -1.82$ ($\tilde t_{\mathrm{i}}=-5.0$), and $\tilde k_{y}=0.5$.  
 }
 \label{fig:typical}
\end{figure}

\begin{figure}
 \begin{center}
  \plotone{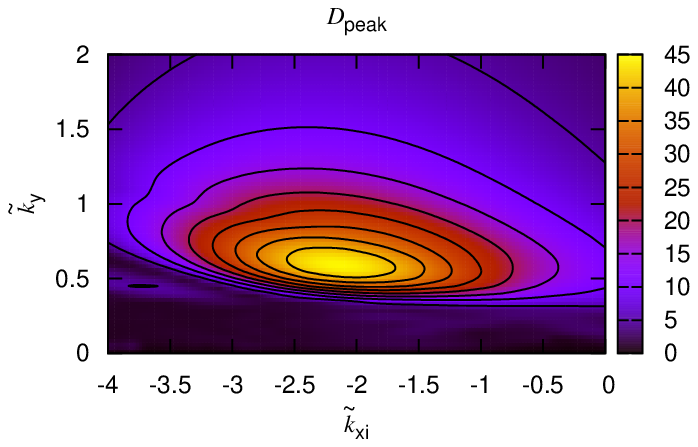}
 \end{center}
 \caption{
 The peak amplitude $D_\mathrm{peak}$ as a function of $\tilde k_{x\mathrm{i}}$
 and $\tilde k_{y}$ for $\tilde \kappa=1.4$ and $Q=1.2$.
 }
 \label{fig:typical2}
\end{figure}

The typical solution is shown in Figure \ref{fig:typical}. 
The parameters are $\tilde \kappa=1.4$, $Q=1.2$, 
 $\tilde k_{x\mathrm{i}}=-1.82$, and $\tilde k_y=0.5$.
The solution has the maximum value $D_{\mathrm{peak}} = 38.8$ at the positive time $\tilde t_\mathrm{peak} = 6.10$, which means that the
 wavelet is trailing when the wavelet is most amplified. 
The peak amplitude $D_{\mathrm{peak}}$ sensitively depends on the
 wavelet.
In the case where we fix $\tilde \kappa$ and $Q$, the peak amplitude has
 the maximum value $D_{\mathrm{max}} $ at $\tilde t_ \mathrm{max}$ for
 $\tilde k_{x\mathrm{i}}= \tilde k_{x\mathrm{i},\mathrm{max}}$ and
 $\tilde k_{y} = \tilde k_{y,\mathrm{max}}$. 
Figure \ref{fig:typical2} shows the dependence of
 $D_{\mathrm{peak}}$ on $\tilde k_{x\mathrm{i}}$ and $\tilde k_y$. 
For $\tilde \kappa=1.4$ and $Q=1.2$, the maximum amplitude is
 $D_\mathrm{max} = 44.8$ at $\tilde t_\mathrm{max} = 5.65$ for 
 $\tilde k_{x\mathrm{i},\mathrm{max}} = -2.2$ and 
 $\tilde k_{y,\mathrm{max}} = 0.60$. 

We assume that wavelets with any wavenumbers always exist because of
 the density fluctuation.  
The particular wavelet with $\tilde k_{x\mathrm{i},\mathrm{max}}$ and
  $\tilde k_{y,\mathrm{max}}$ is amplified most extensively. 
Its amplitude becomes $D_\mathrm{max}$ times larger than the initial amplitude at the positive time $\tilde t_\mathrm{max}$.  
We interpret the most amplified wavelet as the spiral structures
 observed in the simulation. 
The corresponding pitch angle is calculated from $\tilde t_\mathrm{max}$.
From Equation (\ref{eq:pitc}), $\tilde t_\mathrm{max}$ is related to the
 pitch angle: 
\begin{equation}
 \tan \theta_\mathrm{max} = \frac{\tilde \kappa}{ \Gamma \tilde t_\mathrm{max}}.
\label{eq:estb}
\end{equation}

\begin{figure}
 \begin{center}
  \plottwo{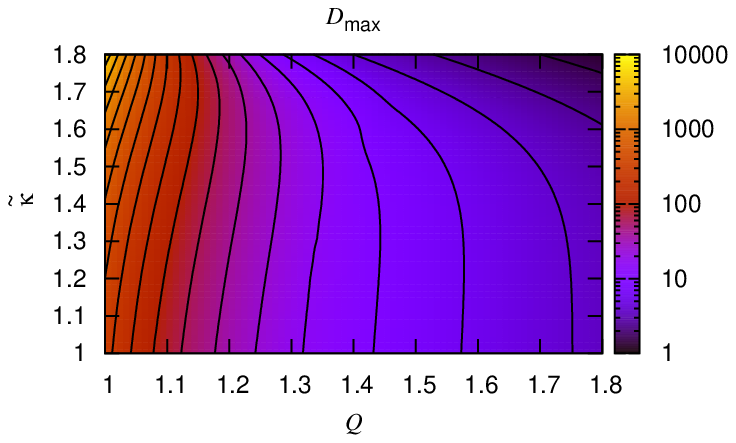}{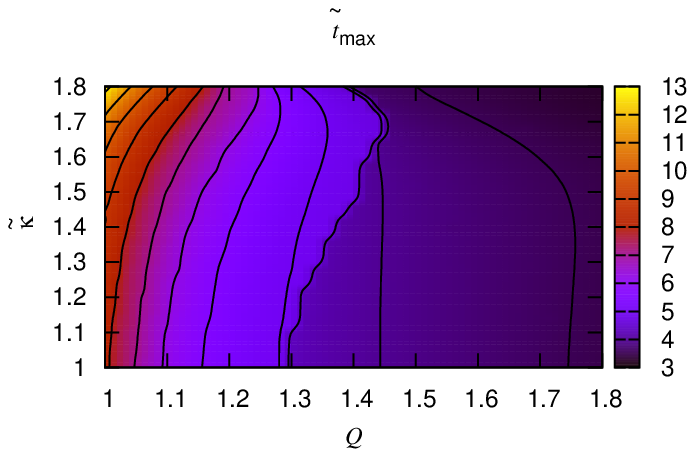}
 \end{center}
 \caption{
The maximum amplitude $D_\mathrm{max}$ (left panel) and the
 corresponding time $\tilde t_\mathrm{max}$ (right panel) as a function
 of $\tilde \kappa$ and $Q$.
}
 \label{fig:tmax}
\end{figure}

The values $D_\mathrm{max}$ and $\tilde t_{\mathrm{max}}$ depend on the
 disk parameters $\tilde \kappa$ and $Q$. 
Figure \ref{fig:tmax} shows $\tilde t_\mathrm{max}$ and $D_\mathrm{max}$
 as a function of $\tilde \kappa$ and $Q$.
The maximum amplitude $D_\mathrm{max}$ depends on $Q$ sensitively.
This is consistent with the results of the $N$-body simulations.
The right panel of Figure \ref{fig:tmax} shows $\tilde t_{\mathrm{max}}$,
 where $\tilde t_\mathrm{max}$ slightly decreases with $Q$ and is
 roughly constant value $\simeq 3.5$ for $Q \gtrsim 1.5$.  

\begin{figure}
 \begin{center}
   \plotone{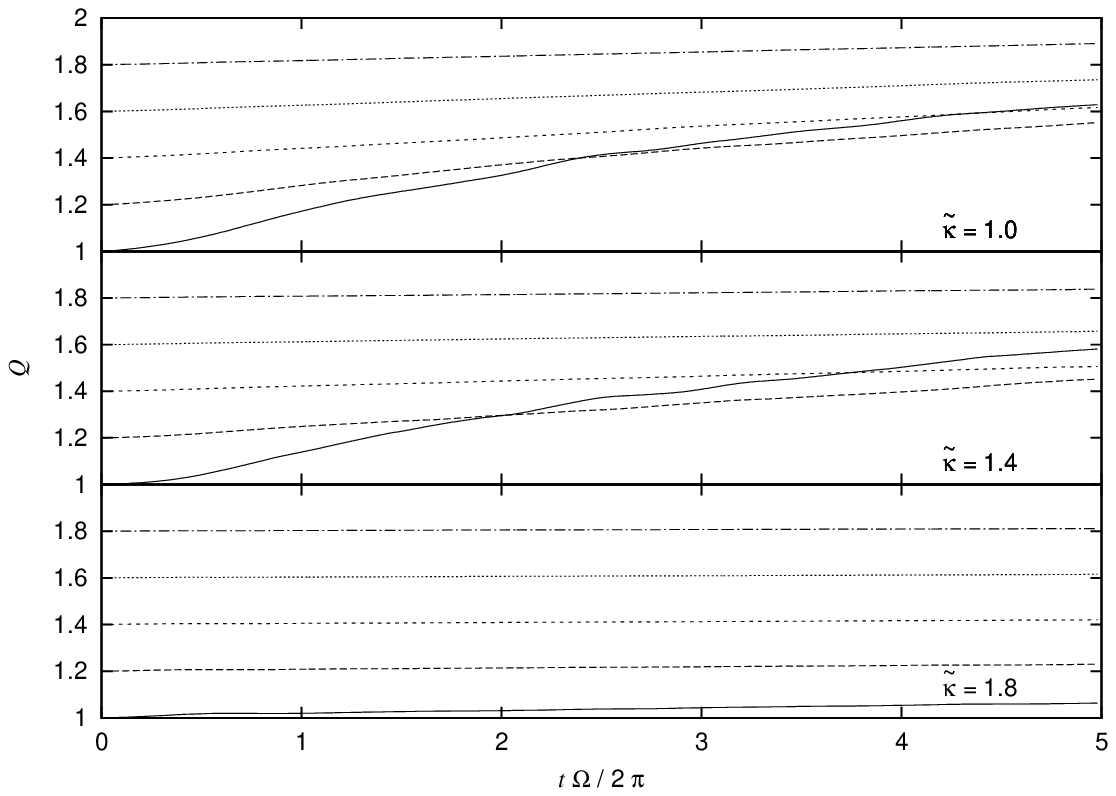}
 \end{center}
 \caption{
The time evolution of Toomre's $Q$ value for the epicycle frequency
 $\tilde \kappa=1.0$ (top), $\tilde \kappa=1.4$ (middle), $\tilde \kappa=1.8$ (bottom) . 
The initial $Q$ value is $1.0$ (solid), $1.2$ (dashed), $1.4$
 (short-dashed), $1.6$ (dotted), and $1.8$ (dot-dashed). 
}
 \label{fig:qval}
\end{figure}

Since $Q$ changes with time in the simulations, we cannot use $Q_\mathrm{ini}$ to calculate the pitch angle.  
Figure \ref{fig:qval} shows the time evolution of $Q$.
The $Q$ value increases more rapidly for smaller $\tilde \kappa$ and initial $Q$ value.
Thus, for $\kappa \lesssim 1.4$, although the initial $Q$ is less than $1.5$, 
the final $Q$ becomes $1.5$ - $2.0$.
Therefore in estimating the pitch angle, we can assume $Q>1.5$ independent of $Q_\mathrm{ini}$. 
As discussed above, $\tilde t_\mathrm{max}$ is roughly constant 
 $\simeq 3.5$ independent of  $\tilde \kappa$ and $Q$ for $Q>1.5$.
Thus, from Equation (\ref{eq:estb}), the pitch angle is estimated as
\begin{equation}
 \tan \theta_\mathrm{max} = \frac{\kappa}{6.9 A},
\label{eq:est2}
\end{equation}
 which agrees well with the fitting formula obtained from the numerical simulations, Equation (\ref{eq:esti}). 

Strictly speaking, for $Q_\mathrm{ini} < 1.4$ and $\tilde \kappa \gtrsim 1.6$ ($\Gamma \lesssim 0.5$), we cannot use that $Q \gtrsim 1.5$ and 
Equation (\ref{eq:est2}).
In fact, Equation (\ref{eq:est2}) for large $\kappa$ (small $\Gamma$) has larger error than that for small $\kappa$ (large $\Gamma$).
However, Equation (\ref{eq:est2}) for small $\Gamma$ explain the general trend of the dependence on $\Gamma$.

\section{Conclusion\label{sec:summary}}

We performed the local $N$-body simulations of stellar disks and
 calculated the pitch angle $\theta$ of the spiral arms as a function of
 the shear rate $\Gamma$. 
We found that $\theta$ is well fitted by Equation (\ref{eq:esti}), which
 agrees well with the observational results \citep{Seigar2006}. 
The pitch angle $\theta$ decreases with $\Gamma$.
For large $\Gamma$ or small $\kappa$, the winding due to the shear is so
 effective that $\theta$ is small. 

We also calculated the time evolution of the wavelet amplitude using the liner theory \citep{Julian1966}. 
The leading wavelet rotates and is amplified owing to the swing
 amplification mechanism \citep{Toomre1981, Toomre1991}. 
The spiral arm can be interpreted as the wavelet amplified by this
 mechanism. 
We calculated the time when the density amplitude is maximum and
 $\theta$ at that time. 
If Toomre's $Q$ value is larger than $1.5$, $\theta$ is approximately
 given by Equation (\ref{eq:est2}). 
Although the initial $Q$ is small, $Q$ increases rapidly due to heating by the spiral arms and exceeds $1.5$ finally. 
Thus, $\theta$ calculated by the numerical simulations agrees with Equation (\ref{eq:est2}). 
All these results suggest that the spiral arms in this simulation are formed by the swing amplification from the leading wavelet in the density fluctuation. 

The present simulation and linear theory employed the local approximation.
We may directly apply these results to flocculent spiral galaxies.
Strictly speaking, we should not apply these results to grand-design spiral galaxies. 
However, we expect that these results are useful for understanding the basic physics of spiral arms in general.

In recent years, it was found that the nonlinear effect is significant to understand the overall activity of the spiral arms \cite[e.g.,][]{Baba2013, DOnghia2013}
We found that the linear theory can predict the correct pitch angle that is consistent with the numerical simulation.
This indicates that the linear theory is still useful to explain the shape of the spiral arms.
We will investigate the non-linear process of spiral arm formation by gravitational instability in more detail in the future work.

\acknowledgments{
Numerical computations were carried out on GRAPE system at Center for Computational Astrophysics, National Astronomical Observatory of Japan.
}


\begin{thebibliography}{27}
\expandafter\ifx\csname natexlab\endcsname\relax\def\natexlab#1{#1}\fi

\bibitem[{{Baba} {et~al.}(2009){Baba}, {Asaki}, {Makino}, {Miyoshi}, {Saitoh},
  \& {Wada}}]{Baba2009}
{Baba}, J., {Asaki}, Y., {Makino}, J., {Miyoshi}, M., {Saitoh}, T.~R., \&
  {Wada}, K. 2009, \apj, 706, 471

\bibitem[{{Baba} {et~al.}(2013){Baba}, {Saitoh}, \& {Wada}}]{Baba2013}
{Baba}, J., {Saitoh}, T.~R., \& {Wada}, K. 2013, \apj, 763, 46

\bibitem[Binney \& Tremaine(2008)]{Binney2008} Binney, J., \& Tremaine, S.\ 2008, Galactic Dynamics: Second Edition, by James Binney and Scott Tremaine.~ISBN 978-0-691-13026-2 (HB).~Published by Princeton University Press, Princeton, NJ USA, 2008.

\bibitem[{{Buta} {et~al.}(2005){Buta}, {Vasylyev}, {Salo}, \&
  {Laurikainen}}]{Buta2005}
{Buta}, R., {Vasylyev}, S., {Salo}, H., \& {Laurikainen}, E. 2005, \aj, 130,
  506

\bibitem[D'Onghia et al.(2013)]{DOnghia2013} D'Onghia, E., Vogelsberger, M., \& Hernquist, L.\ 2013, \apj, 766, 34 

\bibitem[{{French} {et~al.}(2007){French}, {Salo}, {McGhee}, \&
  {Dones}}]{French2007}
{French}, R.~G., {Salo}, H., {McGhee}, C.~A., \& {Dones}, L. 2007, \icarus,
  189, 493

\bibitem[Fuchs(2001)]{Fuchs2001} Fuchs, B.\ 2001, \aap, 368, 107 

\bibitem[{{Fuchs} {et~al.}(2005){Fuchs}, {Dettbarn}, \& {Tsuchiya}}]{Fuchs2005}
{Fuchs}, B., {Dettbarn}, C., \& {Tsuchiya}, T. 2005, \aap, 444, 1

\bibitem[{{Fujii} {et~al.}(2011){Fujii}, {Baba}, {Saitoh}, {Makino}, {Kokubo},
  \& {Wada}}]{Fujii2011}
{Fujii}, M.~S., {Baba}, J., {Saitoh}, T.~R., {Makino}, J., {Kokubo}, E., \&
  {Wada}, K. 2011, \apj, 730, 109

\bibitem[{{Gammie}(2001)}]{Gammie2001}
{Gammie}, C.~F. 2001, \apj, 553, 174

\bibitem[{{Goldreich} \& {Lynden-Bell}(1965)}]{Goldreich1965}
{Goldreich}, P. \& {Lynden-Bell}, D. 1965, \mnras, 130, 125

\bibitem[{{Grand} {et~al.}(2013){Grand}, {Kawata}, \& {Cropper}}]{Grand2013}
{Grand}, R.~J.~J., {Kawata}, D., \& {Cropper}, M. 2013, \aap, 553, A77

\bibitem[Ida et al.(1993)]{Ida1993} Ida, S., Kokubo, E., \& Makino, J.\ 1993, \mnras, 263, 875 

\bibitem[{{Julian} \& {Toomre}(1966)}]{Julian1966}
{Julian}, W.~H. \& {Toomre}, A. 1966, \apj, 146, 810

\bibitem[{{Kawai} \& {Fukushige}(2006)}]{Kawai2006} Kawai, A., \& Fukushige, T. 2006, Proc. 2006 ACM/IEEE Conf. on Supercomputing

\bibitem[{{Kokubo} \& {Ida}(1992)}]{Kokubo1992}
{Kokubo}, E. \& {Ida}, S. 1992, \pasj, 44, 601

\bibitem[{{Lin} \& {Shu}(1964)}]{Lin1964}
{Lin}, C.~C. \& {Shu}, F.~H. 1964, \apj, 140, 646

\bibitem[{{Lin} \& {Shu}(1966)}]{Lin1966} ---. 1966, Proceedings of the National Academy of Science, 55, 229

\bibitem[Oh et al.(2008)]{Oh2008} Oh, S.~H., Kim, W.-T., Lee, H.~M., \& Kim, J.\ 2008, \apj, 683, 94 

\bibitem[{{Salo}(1992)}]{Salo1992a}
{Salo}, H. 1992, \nat, 359, 619

\bibitem[{{Salo}(1995)}]{Salo1995}
---. 1995, Icarus, 117, 287

\bibitem[{{Salo} \& {Schmidt}(2010)}]{Salo2010}
{Salo}, H. \& {Schmidt}, J. 2010, \icarus, 206, 390

\bibitem[{{Seigar} {et~al.}(2005){Seigar}, {Block}, {Puerari}, {Chorney}, \&
  {James}}]{Seigar2005}
{Seigar}, M.~S., {Block}, D.~L., {Puerari}, I., {Chorney}, N.~E., \& {James},
  P.~A. 2005, \mnras, 359, 1065

\bibitem[{{Seigar} {et~al.}(2006){Seigar}, {Bullock}, {Barth}, \&
  {Ho}}]{Seigar2006}
{Seigar}, M.~S., {Bullock}, J.~S., {Barth}, A.~J., \& {Ho}, L.~C. 2006, \apj,
  645, 1012

\bibitem[{{Sellwood}(2000)}]{Sellwood2000}
{Sellwood}, J.~A. 2000, \apss, 272, 31

\bibitem[{{Sellwood}(2010)}]{Sellwood2010}
---. 2010, ArXiv e-prints

\bibitem[{{Sellwood} \& {Carlberg}(1984)}]{Sellwood1984}
{Sellwood}, J.~A. \& {Carlberg}, R.~G. 1984, \apj, 282, 61

\bibitem[Toomre(1964)]{Toomre1964} Toomre, A.\ 1964, \apj, 139, 1217 

\bibitem[Toomre(1981)]{Toomre1981} Toomre, A.\ 1981, Structure and Evolution of Normal Galaxies, 111 

\bibitem[Toomre \& Kalnajs(1991)]{Toomre1991} Toomre, A., \& Kalnajs, A.~J.\ 1991, Dynamics of Disc Galaxies, 341 


\bibitem[{{Toomre} \& {Toomre}(1972)}]{Toomre1972}
{Toomre}, A. \& {Toomre}, J. 1972, \apj, 178, 623

\bibitem[{{Wakita} \& {Sekiya}(2008)}]{Wakita2008}
{Wakita}, S. \& {Sekiya}, M. 2008, \apj, 675, 1559

\bibitem[Wisdom \& Tremaine(1988)]{Wisdom1988} Wisdom, J., \& Tremaine, S.\ 1988, \aj, 95, 925 



\end{thebibliography}
\end{document}